\documentstyle[12pt]{article}

\newcommand{\nc}{\newcommand}
\nc{\beq}{\begin{equation}}
\nc{\eeq}{\end{equation}}
\nc{\beqa}{\begin{eqnarray}}
\nc{\eeqa}{\end{eqnarray}}
\def\DS {D\!\!\!\!/}

\input{epsf.sty}  
\input{psfig.sty}
\newwrite\ffile\global\newcount\figno \global\figno=1

\def\writedef#1{}
\def\figin{\epsfcheck\figin}\def\figins{\epsfcheck\figins}
\def\epsfcheck{\ifx\epsfbox\UnDeFiNeD
\message{(NO epsf.tex, FIGURES WILL BE IGNORED)}
\gdef\figin##1{\vskip2in}\gdef\figins##1{\hskip.5in}
\else\message{(FIGURES WILL BE INCLUDED)}%
\gdef\figin##1{##1}\gdef\figins##1{##1}\fi}
\def\figinsert{}
\def\ifig#1#2#3{\xdef#1{fig.~\the\figno}
\writedef{#1\leftbracket fig.\noexpand~\the\figno}%
\figinsert\figin{\centerline{#3}}\medskip\centerline{\vbox{\baselineskip12pt
\advance\hsize by -1truein\center\footnotesize{  Fig.~\the\figno.} #2}}
\bigskip\endinsert\global\advance\figno by1}
\def\endinsert{}

\begin{document}

\title{\large{\bf Non-Perturbative Couplings and Color Superconductivity}}

\author{ 
Nick Evans\thanks{nevans@budoe.bu.edu}
 \\ {\small Department of Physics, 
Boston University, Boston, MA 02215.} \\ \\
Stephen D.H.~Hsu\thanks{hsu@duende.uoregon.edu}
 \\ {\small Department of Physics, 
University of Oregon, Eugene OR 97403-5203.} \\ \\
Myckola Schwetz\thanks{myckola@baobab.rutgers.edu}
 \\ {\small Department of Physics and Astronomy, 
Rutgers University, 
Piscataway NJ 08855-0849.}\\ \\
}

\date{October  1998 }

\maketitle

\begin{picture}(0,0)(0,0)
\put(350,386){OITS-664}
\put(350,374){RU-98-48}
\put(350,362){BUHEP-98-27}
\end{picture}
\vspace{-24pt}

\begin{abstract}
Quark matter at sufficiently high density exhibits 
color superconductivity, due to attractive gluonic interactions. 
At lower densities of order $\Lambda_{QCD}^3$, it has been proposed 
that instanton generated vertices may play an important role in the 
Cooper pair formation. We study the renormalization 
group flow to the Fermi surface of the full set of couplings generated by 
gluonic and instanton interactions. In earlier work we showed that if the 
gluonic interactions dominate at the matching scale, their running 
determines the scale of the Cooper pair formation $\Delta$.
Here we consider all possibilities, including
the one in which the instanton interactions dominate all others 
at the matching scale. In the latter case we find that a number of 
additional induced couplings 
(including the gluonic ones) reach  their Landau 
poles almost simultaneously with the instanton vertex. Presumably 
all contribute to the Cooper pair formation. The most important
consequence of including all the couplings is a large increase in
the gap size $\Delta$.
\end{abstract}

\newpage

\section{Introduction}

Quark matter at high density exhibits color superconductivity
through the dynamical generation of a Cooper pair 
$\langle \psi^T(-p) C \gamma_5 \psi (p) \rangle$ (see \cite{BL} and
references contained therein). 
A simple way to understand this phenomena is through an effective
field theory description \cite{us,FLRG} of the physics near the Fermi
surface (FS). An attractive coupling between the quarks, such as that 
provided by one gluon exchange, will eventually run to a Landau pole 
as the Wilsonian cutoff approaches the FS. Of course, this is just
a modern reformulation of an insight first gained from the study
of laboratory superconductors \cite{BCSrev}.
At very high densities the effective theory may be
matched to QCD in a naive fashion with perturbative effects like 
one gluon exchange dominating the dynamics \cite{FM}. 
However, when the Fermi momentum is reduced to of order
$\Lambda_{QCD}$ (assuming that this  density is just above the
transition from the low density chiral symmetry breaking phase) it is
possible that additional attractive interactions exist generated by
non-perturbative dynamics. An example of such a vertex is that generated
by instantons (we assume there are two quark flavors throughout
this paper)
\beq
- \kappa  \left( \bar{u}_R T^a  u_L\, \bar{d}_R T^a d_L -
\bar{u}_R T^a d_L \, \bar{d}_R T^a u_L \right)
\eeq
Recent investigations \cite{instanton} examined the role of
this interaction in color superconductivity under the assumption that it
is the unique important coupling for the dynamics. This assumption,
while an interesting starting point for studying the effects of these
non-perturbative couplings, is somewhat ad hoc. Quantum loops generated
by the instanton vertex 
produce additional interactions with different Lorentz structure which,
since the couplings are of order one, we would  expect to be of equal
importance. In this
paper we wish to study the renormalization group flow of the full set of
possible couplings that close under renormalization to determine which
couplings are important to the Cooper pair formation.

QCD at finite density may be described
by the following lagrangian with chemical potential $\mu$
\beq \label{lagrangian}
{\cal L } = - \frac{1}{4}F^{a \mu \nu} F^a_{\mu \nu} ~+~ 
\bar{\psi}_i ( i \DS + \mu  \gamma_0 ) \psi_i~~.
\eeq
We make a guess as to the form of the
effective theory close to the Fermi surface; the obvious guess based on
the dynamics of non-relativistic systems is that the theory is one of
weakly interacting quarks: these are the dressed ``quasi-particles''
of solid state physics language.
Rather than treating the gluons and instantons as dynamical degrees of
freedom we will integrate them out leaving a
potentially infinite sum over higher dimension fermion operators. 
The locality of these operators requires that gluons be screened at 
long-distances, presumably by effective electric and magnetic mass terms 
induced by the medium. 
An electric mass for the gluons arises at leading order in perturbation 
theory, and although there does not appear to be a corresponding magnetic mass 
(at least in perturbation theory), 
Landau damping is sufficient to prevent IR 
divergences due to color-magnetic fluctuations. In the region in which we are 
interested, where instanton effects are important, it seems likely that 
non-perturbative effects (for example, screening due to 
color-magnetic monopole 
excitations) will generate an effective magnetic mass.
A more complete calculation would take into account possible momentum form 
factors in the operators, although it can easily be shown that the RG flows of 
components with different angular momenta decouple near the Fermi surface. For 
simplicity, we will therefore assume that the QCD interactions can be matched 
to momentum independent (s-wave) four fermion operators.
We assume
that the typical gauge propagator has momentum of order $\mu$. Since the
Fermi surface breaks the $O(3,1)$ invariance of the theory to $O(3)$ we
must treat spatial and temporal interactions independently. The full
set of couplings we consider are (up to parity transformations)
\beq \label{couple}
\begin{array}{l}
 g_1 ~ ( \bar{u}_L\, \gamma_0  u_L ~ \bar{u}_L\, \gamma_0  u_L   ~+~
     u ~\Leftrightarrow~ d) \\
 g_2 ~ ( \bar{u}_L\, \gamma_i  u_L ~ \bar{u}_L\, \gamma_i  u_L   ~+~
     u ~\Leftrightarrow~ d) \\
\\
 g_3 ~ ( \bar{u}_L\, \gamma_0  u_L ~ \bar{u}_R\, \gamma_0  \, u_R  ~+~ 
      u \Leftrightarrow d)\\
 g_4 ~ ( \bar{u}_L\, \gamma_i  u_L ~ \bar{u}_R\, \gamma_i  \, u_R  ~+~ 
      u \Leftrightarrow d)\\
\\
  g_5 ~ \bar{u}_L\, \gamma_0  u_L ~ \bar{d}_R\, \gamma_0  d_R  \\
  g_6 ~ \bar{u}_L\, \gamma_i  u_L ~ \bar{d}_R\, \gamma_i  d_R \\
\\
  g_7 ~ \bar{u}_L\, \gamma_0  u_L ~ \bar{d}_L\, \gamma_0  d_L \\
  g_8 ~ \bar{u}_L\, \gamma_i  u_L ~ \bar{d}_L\, \gamma_i  d_L \\
 g_9 ~ \bar{u}_L\, \gamma_0  d_L ~ \bar{d}_L \gamma_0  u_L \\
 g_{10} ~ \bar{u}_L\, \gamma_i  d_L ~ \bar{d}_L \gamma_i  u_L \\
  g_{11} ~ \bar{u}_L  u_R ~ \bar{d}_L  d_R   \\
  g_{12} ~ \bar{u}_L \gamma_0 \gamma_i  u_R ~  \bar{d}_L \gamma_0 
                           \gamma_i  d_R \\
  g_{13} ~ \bar{u}_L  d_R ~ \bar{d}_L  u_R   \\
  g_{14} ~ \bar{u}_L \gamma_0 \gamma_i  d_R ~ 
            \bar{d}_L \gamma_0 \gamma_i  u_R 
\end{array}
\eeq

\bigskip

\noindent 
where the indicated groupings of couplings imply closure under
RG flow. The first three groupings of couplings do not include
all possible gamma matrix structures but we neglect those not present
because the instanton vertex of (1) does not mix under RG flow 
with the one gluon
vertices in these channels. The fourth grouping includes one gluon 
exchange in the left-left channel between two quarks of different 
flavors and the instanton vertex (1). RG closure of the fourth 
grouping requires 
that we include all possible gamma matrix structures.
Henceforth we will include signs from the contraction of spacelike
$\gamma_i$ matrices in the coupling constants.

Here we will consider the $\bar{3}$ channel (which has been shown to be
the most attractive channel for both the gluonic \cite{BL,us}
and instanton \cite{instanton} vertices)
with the color group
matrix structure
\beq
\delta_{ca} \delta_{db} - \delta_{cb} \delta_{da}
\eeq
At high densities, where single gluon exchange dominates the
perturbative effects, the appropriate matching conditions are
\beq \begin{array}{ccc}
g_1 = g_3 = g_5 = g_7 = g = {1 \over 3} 
{ 4 \pi \alpha_s( \mu ) \over \mu^2} 
&& g_2 = g_4 = g_6 = g_8 = - g \\
g_{13} = - g_{11} = \kappa && g_9 = g_{10} = g_{12} = g_{14} = 0 
\end{array}
\eeq
where $\kappa$ is the instanton-generated four-fermion coupling considered
in \cite{instanton}. The
sign of $\kappa$ is a result of the non-perturbative dynamics and is
traditionally \cite{instanton} 
taken positive so (1) is capable of driving chiral
symmetry breaking at $\mu = 0$. Since we are addressing the consistency
of \cite{instanton} we will do likewise.

The Fermi surface in (\ref{lagrangian})  picks out momenta of
order $\mu$. It is therefore natural to study the theory as we approach
the Fermi surface in a Wilsonian sense. We parameterize four momenta in the
following fashion
\beq 
p^{\mu} = (p_0, \vec{p}) = (k_0, \vec{k} + \vec{l})
\eeq
where $\vec{k}$ lies on the Fermi surface and $\vec{l}$ is perpendicular
to it. 
We study the Wilsonian effective theory of modes near the Fermi surface,
with energy and momenta  
\beq
\label{FSE}
|k_0|,|\vec{l}| ~<~ \Lambda~~~~,~~~~ \Lambda \rightarrow 0~~~.
\eeq  
In this limit the four fermion operators are all irrelevant operators
excepting those with the particular three momentum structure
corresponding to quarks with momenta $\vec{k}$ and $ -\vec{k}$ scattering
to momenta $\vec{q}$ and $-\vec{q}$ \cite{us,FLRG}.

As a result of this truncation of the theory 
the only diagrams allowed by the
momentum structure are the bubble diagrams 
found at large N in the familiar O(N) model\footnote{The expansion parameter 
analogous to $1/N$ is $\Lambda/\mu$ \cite{FLRG}.}.  To display
the basic behavior, consider a theory with just the simple interaction
\beq
 G \bar{\psi} \psi \bar{\psi} \psi~~~.
\eeq
The interaction generates
cooper pair formation through the exact gap equation (see Bailin and
Love \cite{BL})
\beq
\Delta = iG \int_0^{\Lambda_{UV}} {d^4p \over (2 \pi)^4 } \Gamma^\mu C(p)
\Gamma_\mu ~~~,
\eeq
where G is the coupling, $\Gamma^\mu$ any associated Dirac 
structure and $C(k)$ a $4\times 4$ off diagonal component of the
$8\times 8$ propagator associated with the fermion vector $(\psi,
\psi^C)$ 
\beq
C(p) = {1 \over ( p\!\!\!/  - \mu \gamma_0)} ~ \Delta ~ {1 \over 
[ \tilde{\Delta} ( p\!\!\!/  - \mu
\gamma_0)^{-1} \Delta - ( p\!\!\!/ + \mu \gamma_0)] }~~~.
\eeq
The gap integral would be log divergent near the FS were it not for the
condensate $\Delta$, which cuts off the contribution of modes with very
low energies. Thus no matter how small the (attractive) 
coupling $G$ there will always be a solution for some $\Delta$. 
Approximating the IR cutoff, and keeping the lowest
order in $\Delta$ we have 
\beq
\Delta  \sim  -i G \int^{\Lambda_{UV}}_\Delta {d^4p \over (2 \pi)^4 } {1
  \over (p\!\!\!/ - \mu \gamma_0)} \Delta {1 \over (p\!\!\!/  + \mu \gamma_0)} 
~~,
\eeq
Neglecting factors from the gamma matrix structure, 
the Cooper pair condensate is of the form
\beq
\Delta = \Lambda_{UV}~ e^{- {c  \over N G}}~~,
\eeq
where $c$ is a constant.

An alternative understanding of this condensate formation is found by
resumming bubble graphs to calculate the renormalization group flow of
the four quark vertex. Note that because of the restricted momentum
structure of the vertices one loop $\beta$ functions are exact.
The one loop graph contributes a logarithmic divergence to the 
running in the presence of a Fermi surface:
\beq
- G^2 \int {d^4p \over (2 \pi)^4} \left[{i \over p^\mu\gamma_\mu + 
\mu \gamma_0 - i \epsilon} \right]_{ik} \left[
 {i \over -p^{\nu} \gamma_\nu + \mu \gamma_0 - i \epsilon} \right]_{jl}
\eeq
Performing the gamma matrix algebra and taking
the limit $k_0, |\vec{l}| \rightarrow 0, 
|\vec{k}|^2 \rightarrow \mu^2$, this
becomes
\beq
\label{gamma}
G^2  \left[  - (\gamma_0)_{ij} (\gamma_0)_{kl} + {1 \over 3}
(\gamma_\alpha )_{ij} (\gamma_\alpha)_{kl}\right]  I
\eeq
where the log divergent part of the integral I is given by
\begin{eqnarray}
I & = &  {1\over 4} 
 \int {dk_0~ d^2k~ dl \over (2 \pi)^4} {1 \over (k_0+l-i \epsilon)(k_0-l
+ i \epsilon)} \nonumber \\
 & \simeq & {i \over 4 } N  \ln \left( {\Lambda_{IR} \over \Lambda_{UV}}
\right)~~~.
\end{eqnarray} 
$N = \int d^2k / (2 \pi)^3 = \mu^2 / 2 \pi^2$, assuming the density of
states at the Fermi surface is given by the lowest order approximation.
The running effective coupling (neglecting the gamma matrix structure
again) is given by
\beq
G(\Lambda_{IR}) = { G({\Lambda_{UV}}) \over 1 + {1 \over 4}
{G(\Lambda_{UV})
N} t }~~~,
\eeq
where $t = \ln ( \Lambda_{IR} / \Lambda_{UV} )$.
Here we have moved from an effective theory with cutoff $\Lambda_{UV}$
($k_0 , |\vec{l}| <  \Lambda_{UV}$), 
to a new effective theory with cutoff $\Lambda_{IR}$.
As we approach the Fermi surface ($\Lambda_{IR}
\rightarrow 0$), the coupling $G$ runs logarithmically. The Landau pole
of the coupling corresponds to the scale $\Delta$ of Cooper pair formation.

In this fashion we can calculate the one loop beta functions for the
vertices in (\ref{couple}) as a function of t. There are in principle $14^2$
entries to the RG matrix but many entries are trivially zero. The 
RG equations simplify drastically if one introduces the linear combinations of 
couplings
\beq
\begin{tabular}{lll}
$G_1 = g_1 + g_2 $ ~~~~  & ~~~~ $G_7 = g_7 + g_8$ ~~~~ & ~~~~ 
  $ G_{11} = g_{11} + g_{12}$  \\
$G_2 = g_1 - 3g_2$ ~~~~ & ~~~~ $G_8 = g_7 - 3g_8 $ ~~~~ & ~~~~
  $ G_{12} = g_{11} - 3g_{12}$ \\
$G_3 = g_3 - g_4 $ ~~~~ & ~~~~ $G_9 = g_9 + g_{10}$ ~~~~ & ~~~~
  $ G_{13} = g_{13} + g_{14}$ \\
$G_4 = g_3 + 3g_4$ ~~~~ & ~~~~ $G_{10} = g_9 - 3g_{10}$ ~~~~ & ~~~~
  $ G_{14} = g_{13} - 3g_{14}$ \\
$G_5 = g_5 - g_6 $ ~~~~  &                        &   \\
$G_6 = g_5 + 3g_6$ ~~~~  &                        &   \\
\end{tabular}
\eeq
and the RG equations then simplify to
\beq
\begin{tabular}{lll}
$\dot{G}_1 = - {1\over 3} N G_1^2$ & 
  $\dot{G}_7 = - {1\over 3} N \Big(G_7^2 + G_9^2 + G_{11}^2 + G_{13}^2\Big)$&  
    $\dot{G}_{11} = - {2 \over 3} N \Big(G_7 G_{11} + G_9 G_{13}\Big) $ \\
$\dot{G}_2 = - N  G_2^2$ &
  $\dot{G}_8 = - N \Big(G_8^2 + G_{10}^2 + G_{12}^2 + G_{14}^2\Big)$& 
    $\dot{G}_{12} = - 2 N  \Big(G_8 G_{12} + G_{10} G_{14}\Big)$ \\ 
$\dot{G}_3 = - {2 \over 3} N G_3^2$ &
  $\dot{G}_9 = - {2 \over 3} N \Big(G_7 G_9 + G_{11} G_{13}\Big)$&
    $\dot{G}_{13} = - {2 \over 3} N \Big(G_7 G_{13} + G_{9} G_{11}\Big)$ \\
$\dot{G}_4 = \dot{G}_6 = 0$ &
  $\dot{G}_{10} = - 2 N \Big(G_8 G_{10} + G_{12} G_{14}\Big)$&
    $\dot{G}_{14} = - 2 N \Big(G_{10} G_{12} + G_8 G_{14}\Big)$ \\
$\dot{G}_5 = - {2 \over 3} N G_5^2$ & & \\
\end{tabular}
\eeq

We proceed further by numerical solution of the RG equations. At very
high density the instanton vertex is expected to be exponentially
suppressed relative to the one gluon exchange. Neglecting the instanton
vertex removes any flavor dependence from the problem and there are
just four independent couplings, the spatial and temporal couplings between 
two left
handed quarks and between a left and a right handed quark. We show the
RG running in Fig 1. The coupling between two left handed quarks
reaches its Landau pole
first as found in \cite{us}.

We may check the effect of the instanton by including it but with
a smaller value than the one gluon exchange at the matching scale. We
show this case in Figure 2 with $\kappa = 0.01 g$. The gluonic couplings
reach their Landau pole before the instanton vertex becomes large, 
though the instanton vertex is eventually driven to infinity by feedback
from the larger gluon interaction. When we include the full RG flow the
instanton vertex does eventually catch the gluonic coupling but only at
very large values of the coupling $\gg 10$, which increase for smaller
values of the instanton coupling at matching. 
The scale of the Landau pole, and
hence the size of the Cooper pairing gap,  is
determined by the running of the gluonic couplings as expected. 

$\left. \right.$ \hspace{-0.4in}
\ifig\prtbdiag{}
{\epsfxsize12.8truecm\epsfbox{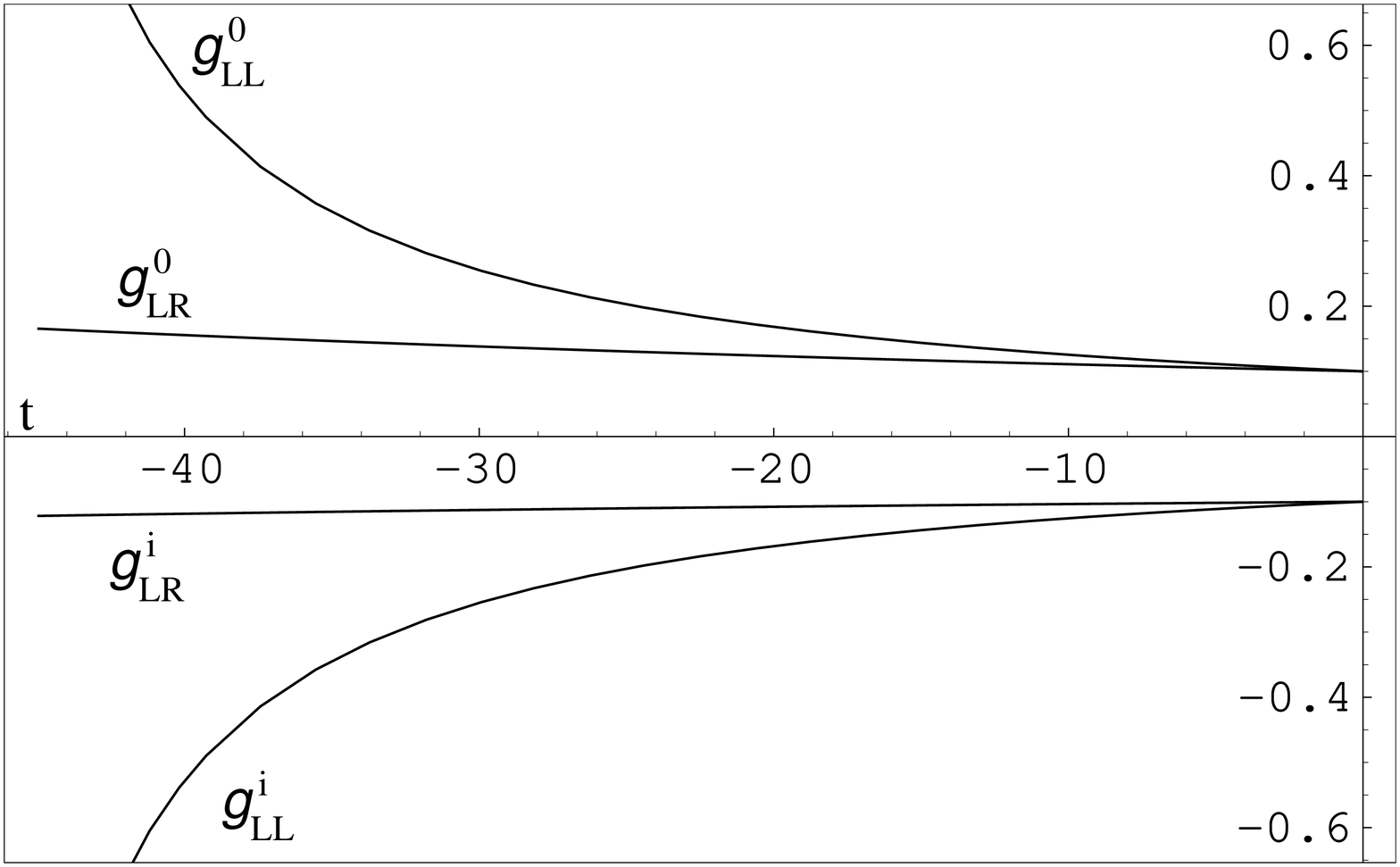}} \vspace{-1cm}
\begin{center} Figure 1:  The RG running of the temporal and spatial one
  gluon derived couplings between two left handed quarks and that
  between a left handed and a right handed quark in the absence of
  instanton effects.
\end{center}

\bigskip

$\left. \right.$ \hspace{-0.4in}
\ifig\prtbdiag{}
{\epsfxsize12.8truecm\epsfbox{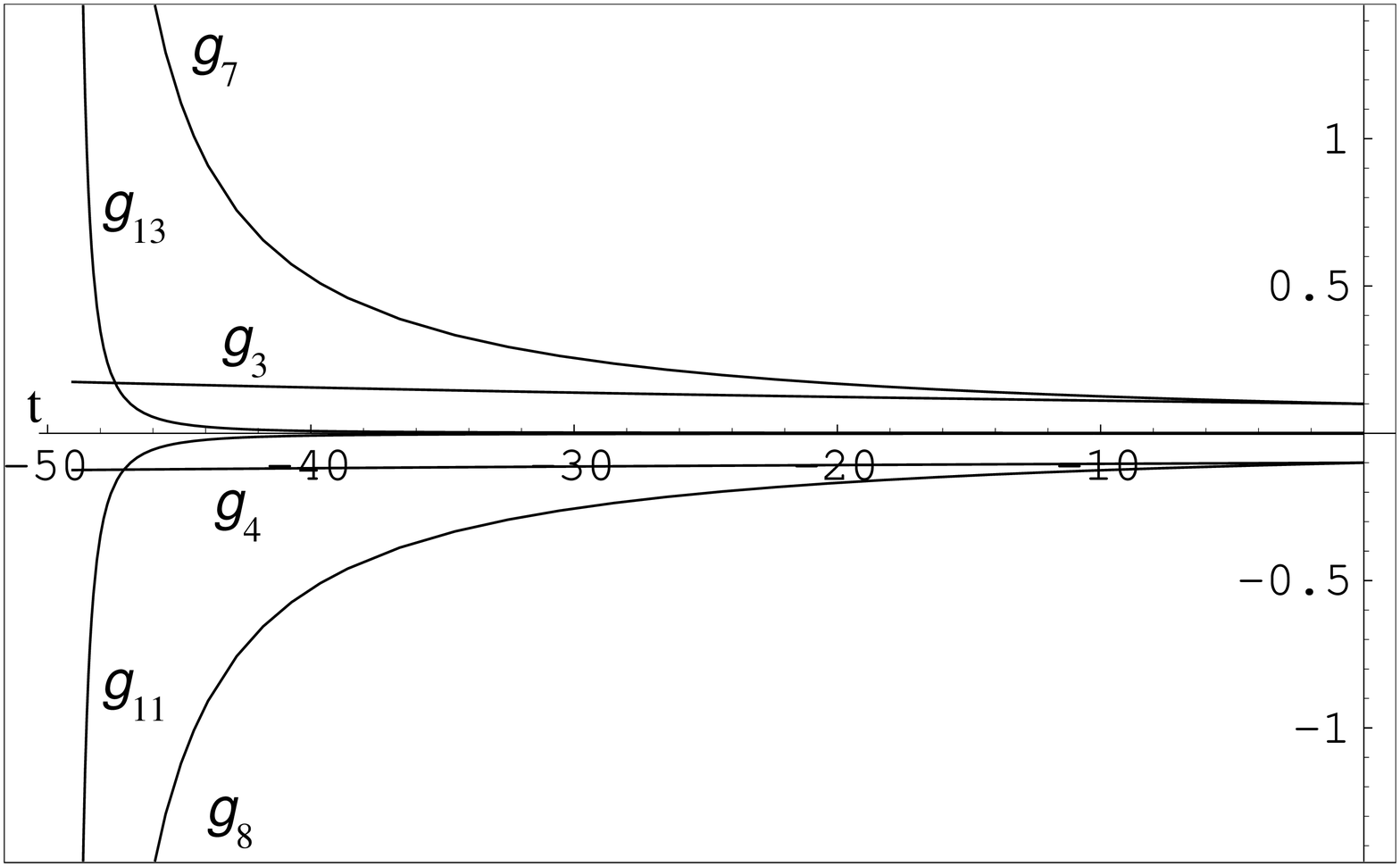}} \vspace{-1cm}
\begin{center} Figure 2: The influence of a small instanton coupling
  ($g_{11}$ and $g_{13}$) on the running of the gluonic couplings
  between an up quark and a down quark ($g_7$ and $g_8$ are in the left
  left channel, $g_3$ and $g_4$ in the left right channel).
\end{center}

We next turn to the effects of the instanton vertex when it is the
dominant interaction. It
is interesting to see what happens in the case where the instanton
vertex is treated as the sole non-zero coupling at the matching
scale. This scenario is shown in Fig 3. The 8 couplings ($g_7 - g_{14}$)
all reach values
of order one simultaneously and presumably play an equal role in Cooper
pair formation. This is not a surprise since, as can be seen from (18), 
the instanton couplings $G_{11}-G_{14}$ only run after 
the generation of couplings $G_7-G_{10}$.
Note that for a very thin shell around the FS
($\Lambda_{UV} << \mu$) this running is implicitly included in a gap
equation analysis performed with just the instanton coupling at the
matching scale, since the gap equation is exact in this limit. 
What this analysis shows is that as expected the 
instanton coupling, which generates the other vertices at one loop,
drives them rather quickly to large values. 
The analysis we have performed retains only the relevant 
operators near the FS. The effects of the intermediate
running (including irrelevant operators) which takes us close to the FS
is assumed to be reflected in our boundary conditions at the 
matching scale. Our results suggest that in QCD at
low densities (of order $\Lambda^3_{QCD}$), where the instanton coupling
$\kappa$ is of order one, the correct analysis would
include all 8 couplings with approximately equal values at matching.

$\left. \right.$ \hspace{-0.4in}
\ifig\prtbdiag{}
{\epsfxsize12.8truecm\epsfbox{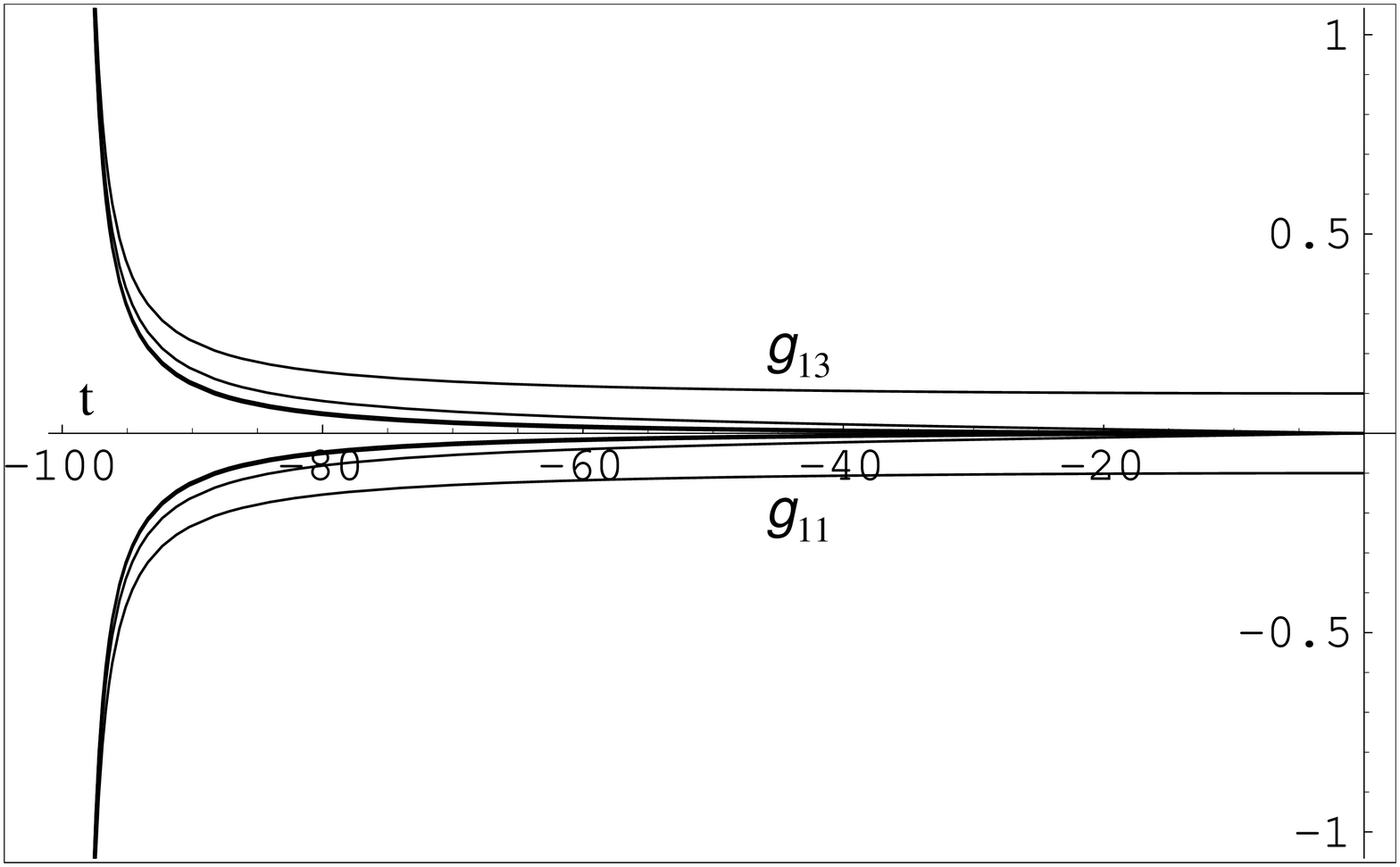}} \vspace{-1cm}
\begin{center} Figure 3: A toy model in which the instanton couplings
  ($g_{11}$ and $g_{13}$) are
  the only non-zero coupling at the matching scale. The other six
  couplings  grow from zero  ($g_7, g_{10}, g_{12}$ have 
  positive values, $g_8, 
  g_{9}, g_{14}$ negative) and
  reach Landau poles at approximately the same scale as the instanton
  vertices. 
\end{center}

If we do assume that all eight interactions are equal at the
matching scale and have the natural choices of sign suggested by Fig 3
then the RG equations simplify. The couplings $G_{\rm odd}$
taken with initial conditions 0 remain 0 for the full running. The
couplings $G_{\rm even}$ with initial conditions $\pm 4G$ run with the
same absolute values and share
the same Landau pole. In terms of the orginal 8 couplings this means
that they all run together and reach their Landau poles
simultaneously. All possible interactions
appear to play an equal role in the Cooper pair formation. We show the
running in Fig 4. 

The Cooper pair formation with only an instanton interaction at matching
has been studied in \cite{instanton} and for the two flavor case the
condensate that forms is in the anti-symmetric color $\bar{3}$, is 
an anti-symmetric singlet in flavor, and an anti-symmetric singlet of
spin (as is the case when the condensate is driven by gluonic
interactions \cite{BL}). 
Since for the instanton case, as can be seen from Fig 3, all the couplings
$g_7-g_{14}$ are essentially equal at the Landau pole we may deduce that
for the case where this equality is enforced at the matching the same
condensate forms. The important difference is that the scale of the
Landau pole is increased sharply by the inclusion of the full set of
couplings at matching. Typically, whatever the matching condition taken on
the couplings, the value of $|t|$ at the pole is significantly
smaller than for the pure instanton case. The Cooper pair condensate, if
non-perturbative effects are sufficiently large to play a role, 
is therefore expected to
be considerably larger than the estimates including only the instanton
vertex.

$\left. \right.$ \hspace{-0.4in}
\ifig\prtbdiag{}
{\epsfxsize12.8truecm\epsfbox{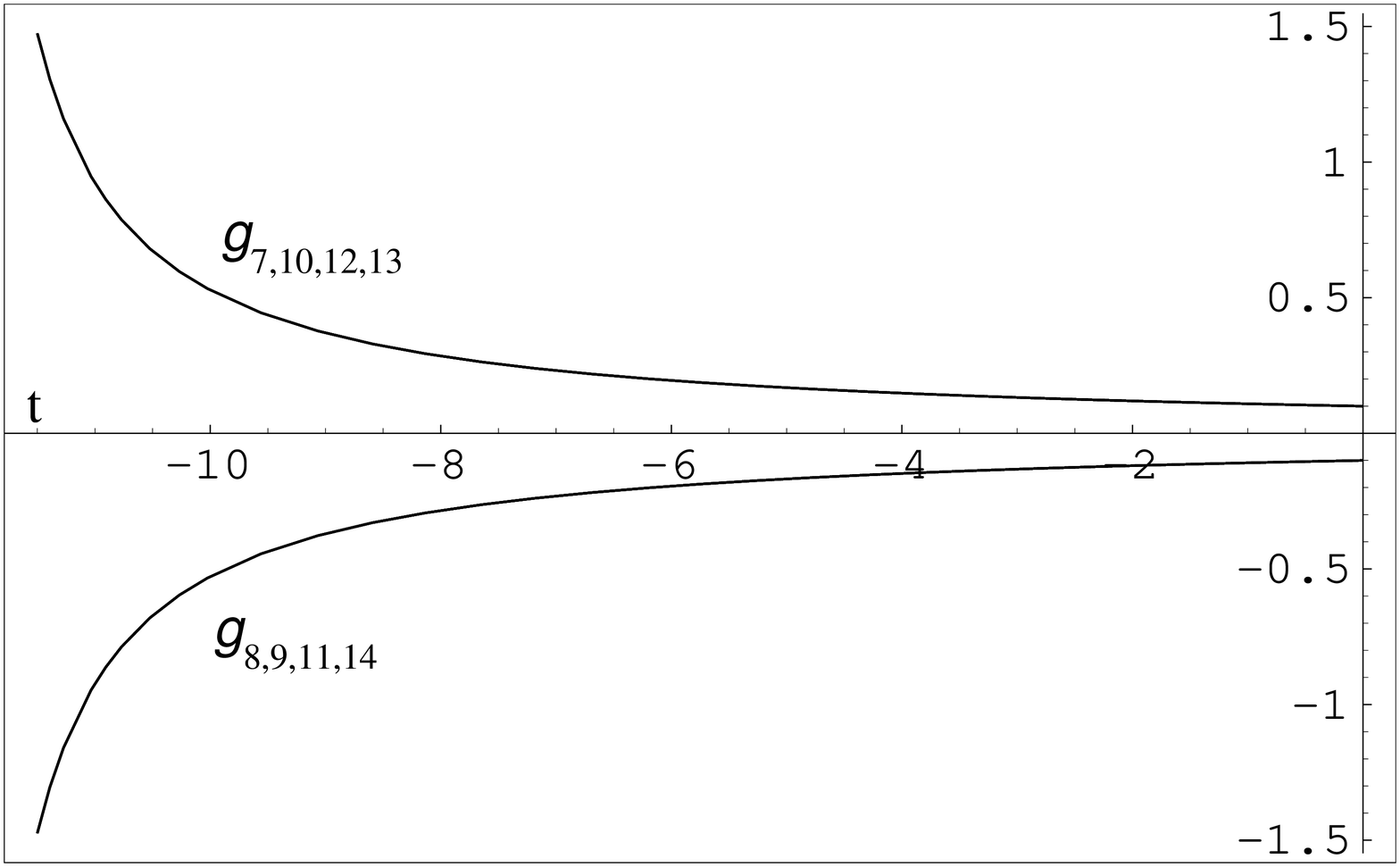}} \vspace{-1cm}
\begin{center} Figure 4: The RG flow when all 8 couplings interacting
  with the instanton vertex ($g_7-g_{14}$) are taken equal at
  matching. The couplings evolve together.
\end{center}
\vspace{1in}

\newpage

\bigskip
\noindent 
The authors would like to thank  M. Alford,  M. Peskin, K. Rajagopal, 
T. Schafer and F. Wilczek for useful discussions and comments. 
After this work was completed we became aware of work by Schafer and
Wilczek (IASNS-HEP-98-90) which addresses similar issues. Their analysis
includes a simplification of the RG equations through the use of flavor
symmetries and Fierz identities.
This work was supported in part under DOE contracts 
DE-FG02-91ER40676, DE-FG06-85ER40224 and DE-FG02-96ER40559


\vskip 1 in
\baselineskip=1.6pt
\begin{thebibliography}{99}
%
%
\def\np#1#2#3{  {Nucl. Phys. #1} (19#3) #2}
\def\pl#1#2#3{  {Phys. Lett. #1} (19#3) #2}
\def\pr#1#2#3{   {Phys. Rev. #1} (19#3) #2}
\def\prep#1#2#3{ {Phys. Rep. #1} (19#3) #2}
\def\prl#1#2#3{ {Phys. Rev. Lett. #1} (19#3) #2}
%


\bibitem{BL} D. Bailin and A. Love, \np{B190}{175}{81};\np{B190}{751}{81};
\np{B205}{119}{82};\prep{107}{325}{84}. 

\bibitem{us} N. Evans, S.D.H. Hsu and M. Schwetz, hep-ph/9808444.

\bibitem{FLRG} G. Benfatto and G. Gallavotti, J. Stat. Phys. 59, 541 (1990);
\pr{C42}{9967}{90}; R. Shankar, Physica A177, 530 (1991); 
Rev. Mod Phys. 66, 129 (1993); J. Polchinski, in Proceedings of the 1992 TASI,
eds. J. Harvey and J. Polchinski (World Scientific, Singapore 1993).

\bibitem{BCSrev} See, for example, N. Ashcroft and N.D. Mermin,
Solid State Physics, Saunders College Publishing (1976).

\bibitem{FM} Freedman and McLerran, \pr{D16}{1130}{77};
\pr{D16}{1147}{77};\pr{D16}{1169}{77}.

\bibitem{instanton}  R. Rapp, T. Schafer, E.V. Shuryak and M. Velkovsky,
 \prl{81}{53}{98};
 M. Alford, K. Rajagopal and F. Wilczek,
\pl{B422}{247}{98}; M. Alford, K. Rajagopal and F. Wilczek, hep-ph/9804403. 



\end{thebibliography}
\end{document}